\documentclass[showpacs,preprint]{revtex4}
\usepackage{graphicx}
\usepackage{subfigure}
\usepackage{amssymb}
\usepackage{amsmath}
\usepackage{bm}
\usepackage{latexsym}
\usepackage{hyperref}

\begin {document}

\title{Quantum dynamics of double-qubits in a spin star lattice with an XY interaction}

\author{Jun Jing$^{1,2}$, Zhi-Guo L\"{u}$^1$\footnote{email:zglv@sjtu.edu.cn}, Hong-Ru. Ma$^{1}$ }
\affiliation{$^1$Department of Physics,
Shanghai Jiaotong University, Shanghai 200240, China \\
$^2$Department of Physics, Shanghai University, Shanghai 200444,
China}

\date{\today}

\begin{abstract}
The dynamics of two coupled spins-1/2 interacting with a spin-bath
via the quantum Heisenberg $XY$ coupling is studied. The pair of
central spins served as a quantum open subsystem are initially
prepared in two types of states: the product states and the Bell
states. The bath, which consists of $N$ (in the thermodynamic limit
$N\rightarrow\infty$) mutually coupled spins-1/2, is in a thermal
state at the beginning. By the Holstein-Primakoff transformation,
the model can be treated effectively as two spin qubits embedded in
a single mode cavity. The time-evolution of the purity, z-component
summation and the concurrence of the central spins can be determined
by a Laguerre polynomial scheme. It is found that (i) at a low
temperature, the uncoupled subsystem in a product state can be
entangled due to the interaction with bath, which is tested by the
Peres-Horodecki separability; however, at a high temperature, the
bath produces a stronger destroy effect on the purity and
entanglement of the subsystem; (ii) when the coupling strength
between the two central spins is large, they are protected strongly
against the bath; (iii) when the interaction between the subsystem
and the bath is strong, the collapse of the two spin qubits from
their initial entangled state is fast.

\end{abstract}
\pacs{75.10.Jm, 03.65.Bz, 03.67.-a}


\maketitle

\section{Introduction}

Compared with other physical systems \cite{Braunstein}, solid-state
devices, in particular, the ultra-small quantum dots \cite{Burkard}
with spin degrees of freedom embedded in nanostructured materials
are more easily scaled up to large registers and they can be
manipulated by energy bias and tunneling potentials \cite{Loss}.
Naturally, the spin systems are very promising candidates for
quantum computation \cite{Loss, Kane, Nielsen}. Inevitably, the spin
qubits are open quantum systems \cite{Breuer2, weiss} subjected to
the interactions with their environments. In a short time, the
states of the qubits will relax into a set of ``pointer states'' in
the Hilbert space \cite{Zurek1981}, which can be quantified by using
the purity \cite{Zurek1987}; the entanglement between the spin
qubits will also vanish. Yet quantum entanglement is the most
intriguing feature of quantum composite system and the vital factor
for quantum computation and quantum communication \cite{Nielsen,
Bennett}. These two disadvantages, so-called decoherence and
disentanglement, will not be overcome until the modelling of the
surrounding environment or bath of the spin systems. \\

For solid state spin nano-devices, the quantum noise, causing
decoherence and disentanglement of the qubits system, mainly arises
from the contribution of nuclear spins, which is usually regarded as
a spin environment. Recently, there are some works devoted to the
behavior of central spins under a strong non-Markovian influence of
a spin-bath \cite{Stamp, Loss2}. Lucamarini et al. had used a
perturbation method \cite{Paganelli} and a mean-field approximation
\cite{Paganelli2} to study the temporal evolution of entanglement
pertaining to two qubits interacting with a thermal bath. They found
entangled states with an exponential decay of the quantum
correlation at finite temperature. Hutton and Bose \cite{Hutton}
investigated a star network of spins, in which all spins interact
exclusively and continuously with a central spin via Heisenberg XX
couplings with equal strength. Their work was advanced by Hamdouni
et al. \cite{Hamdouni}, who derived the exact reduced dynamics of a
central two-qubit subsystem in the same bath configuration. And they
also studied the entanglement evolution of the central system. Yuan
et al. \cite{Yuan} developed a novel operator technique to obtain
the dynamics of the two coupled spins in quantum Heisenberg XY
\cite{Breuer} spin bath with high symmetry. The results of all the
above works are very exciting. Yet all of their methods are of
complex analytical derivations. And in Ref. \cite{Yuan}, their
analytical results are dependent on some particular initial states.
The study of quantum dynamics from different initial states, such as
Bell states and product states, is a very interesting issue in this
system. Thus we introduce a ``half analytical and half numerical''
method here to solve this kind of open quantum problem and show
light on their features of dynamics from different initial states.
Moreover, the numerical part of our method is beyond the Markovian
approximation \cite{Fannes} due to the strong non-Markovian behavior
of such a center-spins-spin-bath model. \\

In this paper, we study an open two-spin-qubit system in a spin
star-like configuration, which is similar to the cases studied in
Ref. \cite{Yuan, Hamdouni}: the interaction among the bath-spins,
between the two qubits and between the subsystem and the bath are
all of the Heisenberg XY type. The present model involves the
Heisenberg XY interaction that has broad applications for various
quantum information processing systems, such as quantum dots,
Cavity-QED, etc\cite{Imamoglu, Zheng, Wang, Lidar}. First, we use
Holstein-Primakoff transformation to reduce the model to an
effective Hamiltonian in the field of cavity quantum electrodynamics
\cite{Milburn}. Second, we apply a numerical simulation to determine
the dynamics of the whole system and obtain the reduced dynamics of
the two coupled spin qubits by tracing over the bath modes. During
our numerical calculation, there are no approximations assumed and
the initial state of the subsystem (a pair of central spins) can be
chosen arbitrarily. We will give some results about the purity, the
z-component summation and the concurrence of the center open spin
subsystem in the thermodynamical limit. Additionally, we will show
that the bath can lead to entanglement of initially unentangled
qubits. The rest of this paper is organized as follows. In Sec.
\ref{Hamiltonian} the model Hamiltonian and the operator
transformation procedure is introduced. In Sec. \ref{calculation},
we explain the numerical techniques about the evolution of the
reduced matrix for the subsystem; Detailed results and discussions
are in Sec. \ref{discussion}; The conclusion of our study is given
in Sec. \ref{conclusion}.

\section{Model and Transformation}\label{Hamiltonian}

Consider a two-spin-qubit subsystem symmetrically interacting with
bath spins via a Heisenberg XY interaction: both the subsystem and
the bath are composed of spin-1/2 atoms. Each spin in the bath
interacts with the two center ones by the same coupling strength,
which is similar to the cases in the literature\cite{Yuan, Breuer,
Hutton, Breuer3}. The interactions between bath spins are also of
the XY type. The Hamiltonian for the total system is
\begin{eqnarray}\label{Hami}
H&=&H_S+H_{SB}+H_B.\\ \label{H_S}
H_S&=&\mu_0(S_{01}^z+S_{02}^z)+\Omega\left(S_{01}^+S_{02}^-+S_{01}^-S_{02}^+\right),\\
\label{HSB}
H_{SB}&=&\frac{g_0}{\sqrt{N}}\left[\left(S_{01}^++S_{02}^+\right)
\sum_{i=1}^N
S_i^-+\left(S_{01}^-+S_{02}^-\right)\sum_{i=1}^NS_i^+\right],\\
H_B&=&\frac{g}{N}\sum_{i\neq j}^N\left(S_i^+S_j^-+S_i^-S_j^+\right).
\end{eqnarray}
Here, $H_S$ and $H_B$ describe the subsystem and bath respectively,
and $H_{SB}$ is the interaction part in the whole Hamiltonian
\cite{Breuer, Yuan, Canosa}. $\mu_0$ represents the coupling
constant between a locally applied external magnetic field along the
$z$ direction and the spin qubit subsystem. $\Omega$ is the coupling
constant between two qubit spins. $S_{0i}^+$ and $S_{0i}^-$
($i$=1,2) are the spin-flip operators of the subsystem qubits,
respectively, which are:
\begin{equation}
S^+=\left(\begin{array}{cc}
      0 & 1 \\
      0 & 0
    \end{array}\right), \quad
S^-=\left(\begin{array}{cc}
      0 & 0\\
      1 & 0
    \end{array}\right).
\end{equation}
$S_i^+$ and $S_i^-$ are the corresponding operators for the $i$th
atom spin in the bath. The index $i$ in the summation runs from $1$
to $N$, where $N$ is the number of the bath spins. $g_0$ is the
coupling constant between the subsystem and the bath, and $g$
is the one between any two spins in the bath.\\

Substituting the collective angular momentum operators
$J_{\pm}=\sum_{i=1}^NS_{i}^\pm$ into Eq. \ref{HSB}, we rewrite the
last two parts of the Hamiltonian as:
\begin{eqnarray}\label{H_SB1}
H_{SB}&=&\frac{g_0}{\sqrt{2j}}\left[\left(S_{01}^++S_{02}^+\right)J_-+
\left(S_{01}^-+S_{02}^-\right)J_+\right],\\ \label{H_B1}
H_B&=&\frac{g}{2j}\left(J_+J_-+J_-J_+\right)-g,
\end{eqnarray}
where $j=N/2$. After the Holstein-Primakoff transformation (It
transforms the spin bath of infinity spins into an effective boson
bath) \cite{Holstein},
\begin{equation}\label{J}
J_+=b^+\left(\sqrt{2j-b^+b}\right), \hspace*{2mm}
J_-=\left(\sqrt{2j-b^+b}\right)b,
\end{equation}
with $[b,b^+]=1$, the Hamiltonian, Eqs. \ref{H_SB1} and \ref{H_B1},
can be written as
\begin{eqnarray}\label{H_SB2}
H_{SB}&=&g_0\left[\left(S_{01}^++S_{02}^+\right)\sqrt{1-\frac{b^+b}{N}}b+
\left(S_{01}^-+S_{02}^-\right) b^+\sqrt{1-\frac{b^+b}{N}}\right],\\
\label{H_B2} H_B&=&g\left[ b^+\left(1-\frac{b^+b}{N}\right)
b+\sqrt{1-\frac{b^+b}{N}}bb^+\sqrt{1-\frac{b^+b}{N}}\right]-g.
\end{eqnarray}
In the thermodynamic limit (i.e. $N\longrightarrow\infty$) at
finite temperatures, we then have
\begin{eqnarray}\label{H_SB}
H_{SB}&=&g_0\left[\left(S_{01}^++S_{02}^+\right)
b+\left(S_{01}^-+S_{02}^-\right) b^+\right],\\ \label{H_B} H_B&=&2g
b^+b.
\end{eqnarray}
Although the whole Hamiltonian composed by Eqs. \ref{H_S},
\ref{H_SB} and \ref{H_B} is similar to that of a Jaynes-Cumming
model \cite{Yuan}, there is an explicit difference between the two
models. The present Hamiltonian describes two coupled qubits
interacting with a single-mode thermal bosonic bath field, so the
analysis of the model is a nontrivial problem in cavity quantum
electrodynamics \cite{Imamoglu, Zheng}. We note here that due to the
transition invariance of the bath spins in our model, it is
effectively represented by a single collective environment
pseudo-spin $J$ in Eq. \ref{J}. After the Holstein-Primakoff
transformation and in the thermodynamic limit, this collective
environment pseudo-spin could be considered a single-mode bosonic
thermal field. The effect of this single-mode environment on the
dynamics of the two coupled qubits is interesting and nontrivial. In
Sec. \ref{discussion}, we will show some results, for example, the
revival behavior of the reduced density matrix and the entanglement
evolution of the two central spins. This could be used in real
quantum information applications.

\section{Numerical Calculation procedures}\label{calculation}

The initial density matrix of the total system is assumed to be
separable, i.e., $\rho(0)=|\psi\rangle\langle\psi|\otimes\rho_B$.
The density matrix of the spin bath satisfies the Boltzmann
distribution, that is $\rho_B=e^{-H_B/T}/Z$, where $Z={\rm
Tr}\left(e^{-H_B/T}\right)$ is the partition function, and the
Boltzmann constant $k_B$ has been set to $1$ for sake of simplicity.
The density matrix $\rho(t)$ of the whole system can be derived by
\begin{eqnarray}
\rho(t)&=&\exp(-iHt)\rho(0)\exp(iHt),\\ \label{rhoB}
\rho(0)&=&\rho_S(0)\otimes\rho_B(0),\\ \label{rho_s}
\rho_S(0)&=&|\psi(0)\rangle\langle\psi(0)|.
\end{eqnarray}
In order to find the density matrix $\rho(t)$, we follow the method
suggested by Tessieri and Wilkie \cite{TWmodel}. The thermal bath
state $\rho_B(0)$ can be expanded with the eigenstates of the
environment Hamiltonian $H_B$:
\begin{eqnarray}\label{rho_B}
\rho_B(0)&=&\sum_{m=1}^M|\phi_m\rangle\omega_m\langle\phi_m|,\\
\label{weight}
\omega_m&=&\frac{e^{-E_m/T}}{Z},\\
Z&=&\sum_{m=1}^Me^{-E_m/T}.
\end{eqnarray}
Here $|\phi_m\rangle$, $m=1, 2, 3, \cdots, M$, are the eigenstates
of $H_B$ and $E_m$ are the corresponding eigen energies. According
to the form of Eq. \ref{H_B}, $|\phi_m\rangle=|m\rangle$ and
$E_m=2gm$, thus $\omega_m=\exp(-2gm/T)$ and $Z=1/(1-e^{-2g/T})$.
With this expansion, the density matrix $\rho(t)$ can be written as:
\begin{equation}\label{equ:2m}
\rho(t)=\sum_{m=1}^M\omega_m|\Psi_m(t)\rangle\langle\Psi_m(t)|.
\end{equation}
Where
\begin{equation}
|\Psi_m(t)\rangle =\exp(-iHt)|\Psi_m(0)\rangle
=U(t)|\Psi_m(0)\rangle.
\end{equation}
The initial state is
\[
|\Psi_m(0)\rangle =|\psi(0)\rangle|m\rangle.
\]
The evolution operator $U(t)$ can be evaluated by different methods.
In Ref. \cite{Yuan}, they use a unique analytical operator
technique, which is dependent on the special initial state. Here, we
apply an efficient numerical algorithm based on polynomial schemes
\cite{Jing, Dobrovitski1, Hu} into this problem. The method used in
this calculation is the Laguerre polynomial expansion method we
proposed in Ref. \cite{Jing}, which is pretty well suited to many
quantum systems and can give accurate result with a comparatively
smaller computation load. More precisely, the evolution operator
$U(t)$ is expanded in terms of the Laguerre polynomial of the
Hamiltonian as:
\begin{equation*}
U(t) =  \left(\frac{1}{1+it}\right)^{\alpha+1}
\sum^{\infty}_{k=0}\left(\frac{it}{1+it}\right)^kL^{\alpha}_k(H),
\end{equation*}
where $\alpha$ distinguishes different types of Laguerre polynomials
\cite{Arfken}; $k$ is the order of the Laguerre polynomial. In
practice, the expansion has to be cut at some value of
$k_{\text{max}}$ for the compromise of the numerical stability in
the recurrence of the Laguerre polynomial and the speed of
calculation. $k_{\text{max}}$ is optimized to be $20$ in this study
and the time step $t$ is restricted to some value in order to get
accurate results of the evolution operator. At every time step, the
accuracy of the results will be confirmed by the test of the
numerical stability --- whether the trace of the density matrix is
$1$ with error less than $10^{-12}$. In a longer time scale, the
evolution can be achieved with more steps. The action of the time
evolution operator is calculated by utilizing recurrence relations
of the Laguerre polynomial. The efficiency of this scheme
\cite{Jing} is about $10$ times as that of the Runge-Kutta algorithm
used in Ref. \cite{TWmodel}. When the states $|\Psi_m(t)\rangle$ are
obtained, the density matrix can be found by performing the
summation in Eq. (\ref{equ:2m}).\\

In principle we should consider every energy level of the
single-mode bath field: $M\rightarrow\infty$. But the contribution
of the high energy states $|m\rangle, m>m_C$ ($m_C$ is a cutoff to
the spin bath eigenstates) is found to be negligible due to their
weights $\omega_m$ as long as the temperature is finite. That is to
say, Eq. (\ref{equ:2m}) could be changed to the following form:
\begin{equation}\label{equ:M}
\rho(t)=\sum_{m=1}^{m_C}\omega_m|\Psi_m(t)\rangle\langle\Psi_m(t)|.
\end{equation}

Given the density matrix of the whole system $\rho(t)$, we can find
the reduced density matrix by a partial trace operation, which
traces out the freedom degrees of the environment:
\begin{equation}\label{final}
\rho_S(t)={\rm Tr}_B\left(\rho(t)\right).
\end{equation}
For our model, $\rho_S=|\psi\rangle\langle\psi|$ is a density matrix
of the open subsystem consists of two coupled central spins, which
can be expressed by a $4\times4$ matrix in the subsystem Hilbert
space spanned by the orthonormal vectors $|00\rangle$, $|01\rangle$,
$|10\rangle$ and $|11\rangle$. The most general form of an initial
pure state of the two-qubit system is
\begin{eqnarray}
|\psi(0)\rangle=\alpha|00\rangle+\beta|11\rangle+\gamma|01\rangle+\delta|10\rangle,\\
\mbox{with}\quad |\alpha|^2+|\beta|^2+|\gamma|^2+|\delta|^2=1.
\end{eqnarray}

\section{Numerical simulation results and discussions}
\label{discussion}

\begin{figure}[htbp]
\centering \subfigure[Purity]{\label{T11:Purity}
\includegraphics[width=3in]{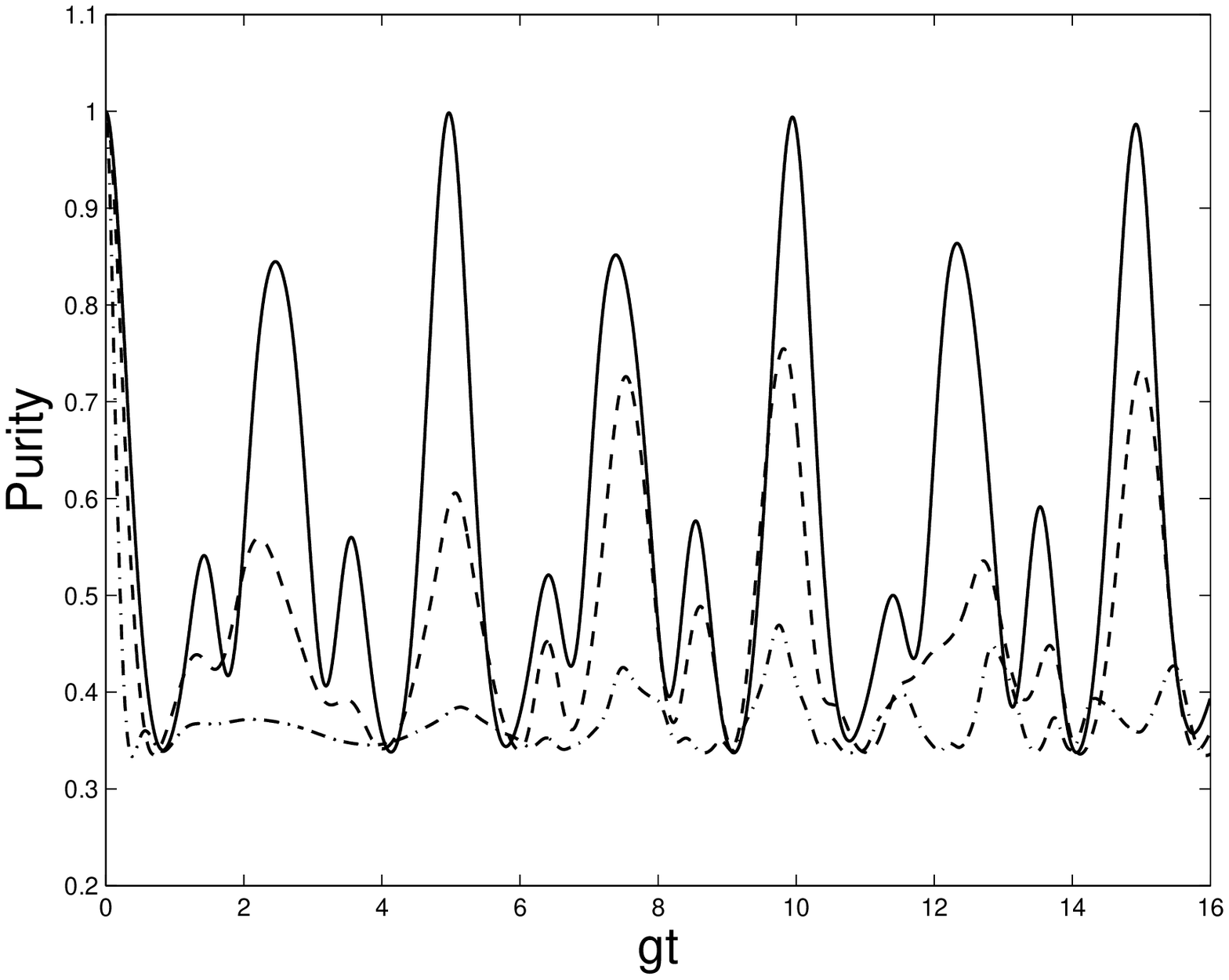}}
\subfigure[$\langle{}S^{z}_{01}+S^{z}_{02}\rangle$]{\label{T11:Sz}
\includegraphics[width=3in]{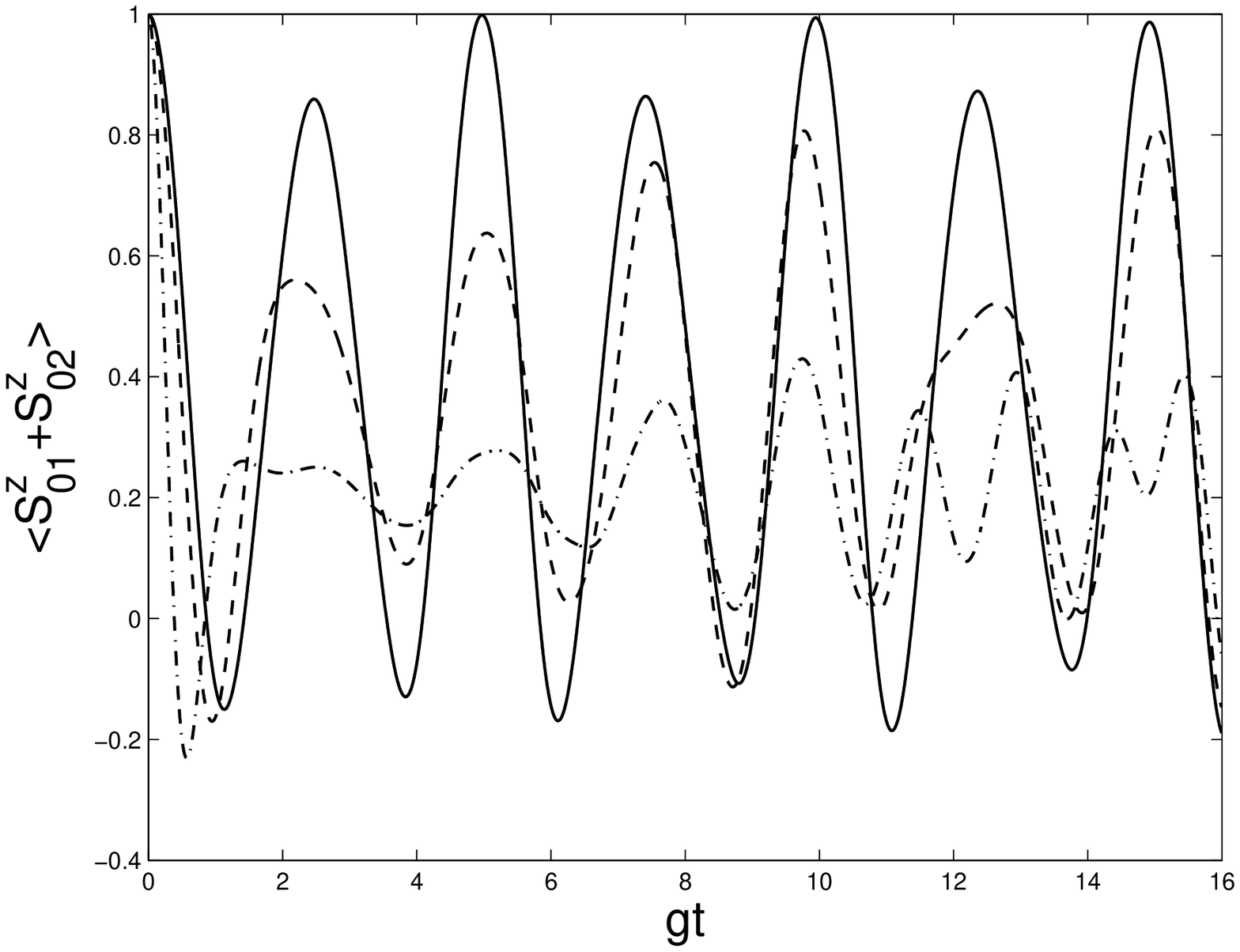}}
\caption{Time evolution of (a) purity and (b) z-component
oscillation for an initial two-qubit state of
$|\psi(0)\rangle=|11\rangle$ for different values temperature:
$T=0.2g$ (solid curve), $T=2g$ (dashed curve) and $T=10g$ (dot
dashed curve). Other parameters are $\mu_0=2g$, $g_0=g$,
$\Omega=0g$.} \label{T11}
\end{figure}

\begin{figure}[htbp]
\centering
\includegraphics[scale=0.6]{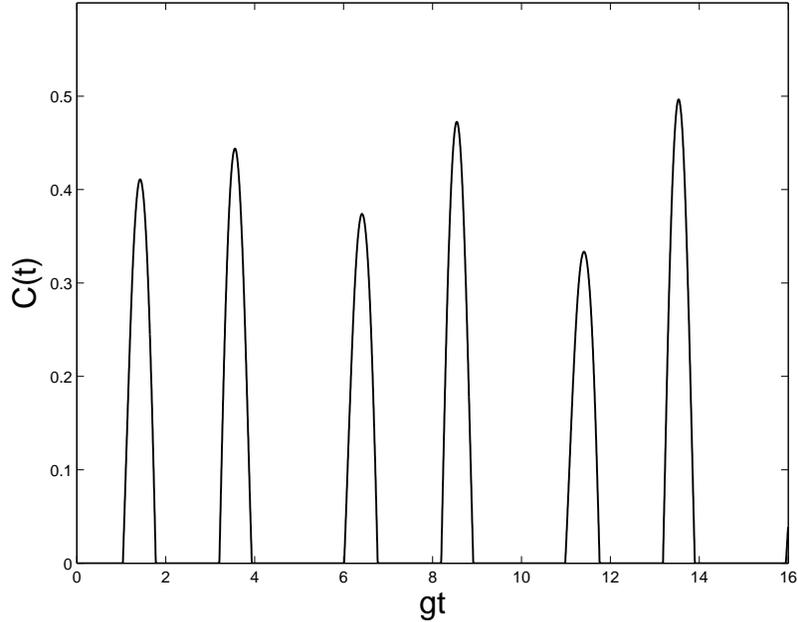}
\caption{Time evolution of the concurrence for an initial
two-qubit state of $|\psi(0)\rangle=|11\rangle$ at low
temperature: $T=0.2g$. Other parameters are $\mu_0=2g$, $g_0=g$,
$\Omega=0g$.} \label{Con-Pu}
\end{figure}

\begin{figure}[htbp]
\centering \subfigure[Purity]{\label{Omega11:Purity}
\includegraphics[width=3in]{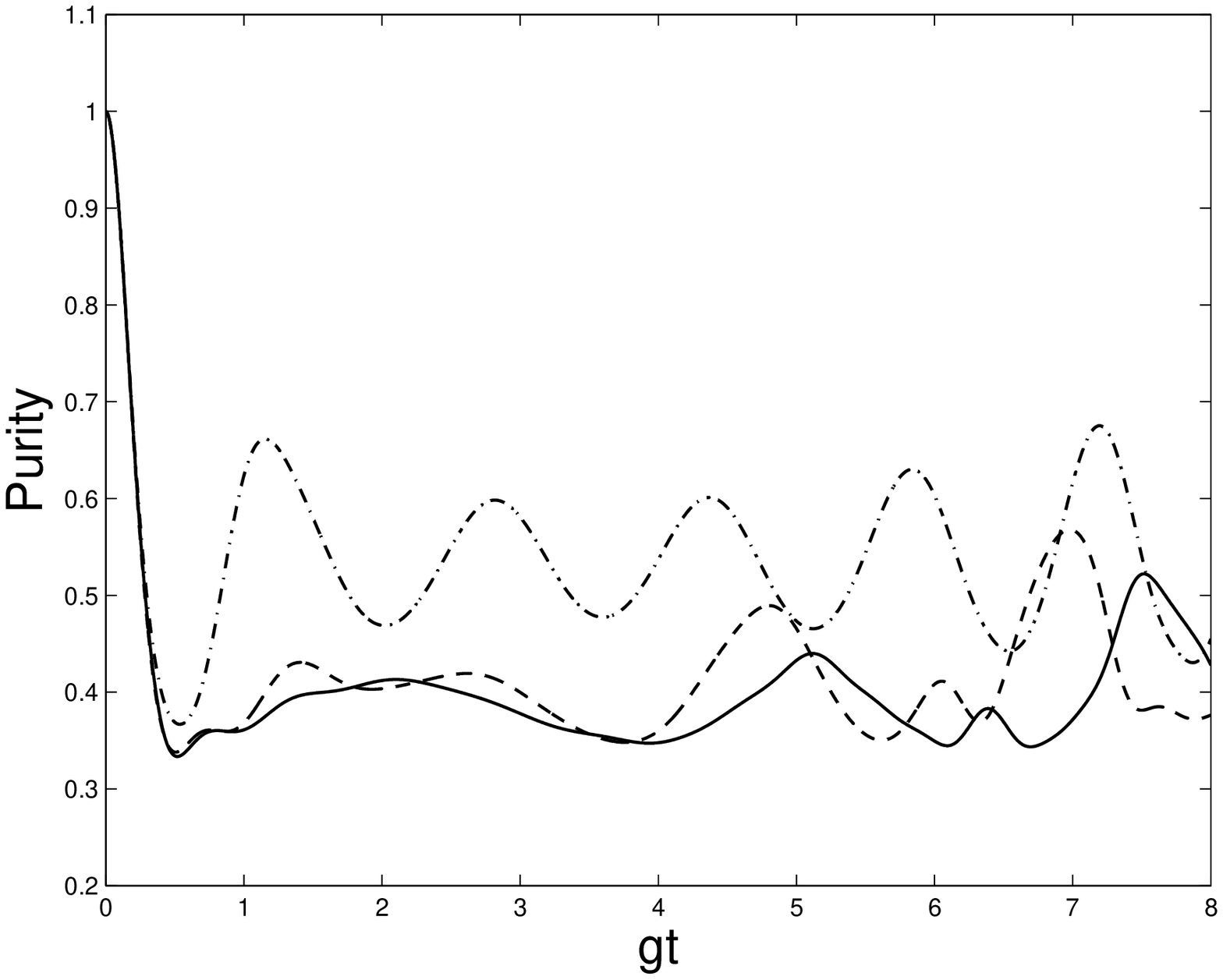}}
\subfigure[$\langle{}S^{z}_{01}+S^{z}_{02}\rangle$]{\label{Omega11:Sz}
\includegraphics[width=3in]{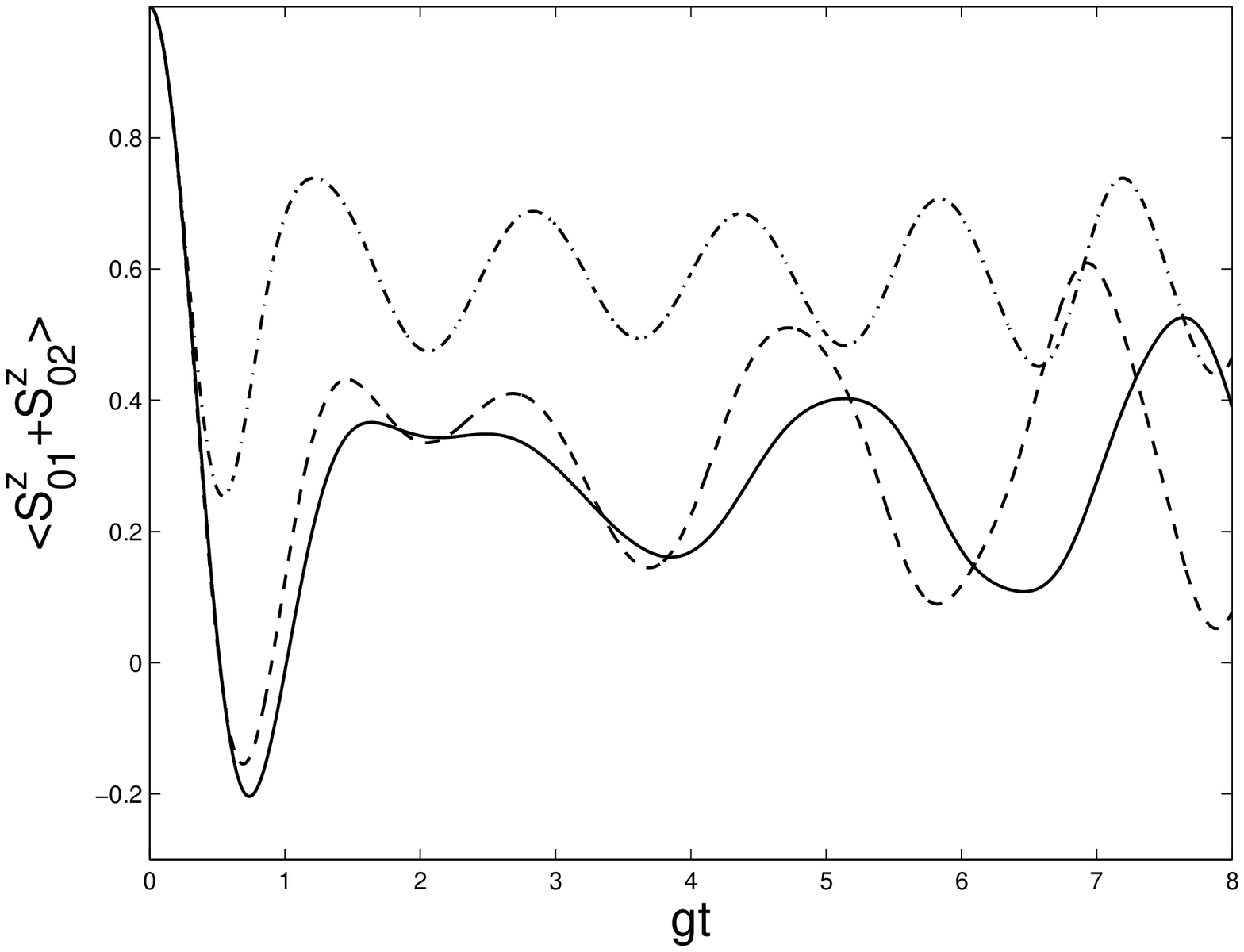}}
\caption{Time evolution of (a) purity and (b) z-component
oscillation for an initial two-qubit state of
$|\psi(0)\rangle=|11\rangle$ for different values $\Omega$:
$\Omega=0g$ (solid curve), $\Omega=2g$ (dashed curve) and
$\Omega=5g$ (dot dashed curve). Other parameters are $\mu_0=2g$,
$g_0=g$, $T=5g$.} \label{Omega11}
\end{figure}

\begin{figure}[htbp]
\centering \subfigure[Purity]{\label{g011:Purity}
\includegraphics[width=3in]{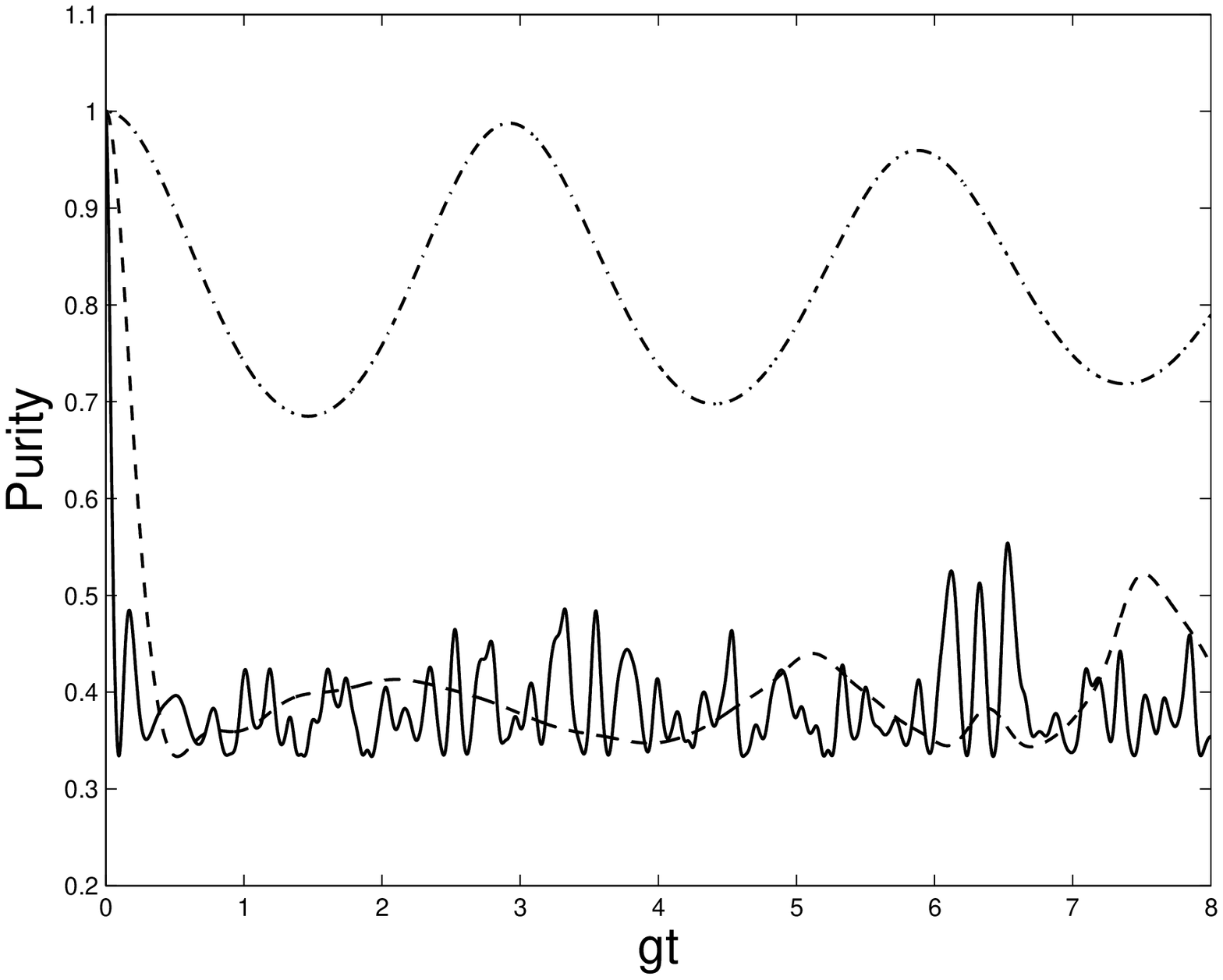}}
\subfigure[$\langle{}S^{z}_{01}+S^{z}_{02}\rangle$]{\label{g011:Sz}
\includegraphics[width=3in]{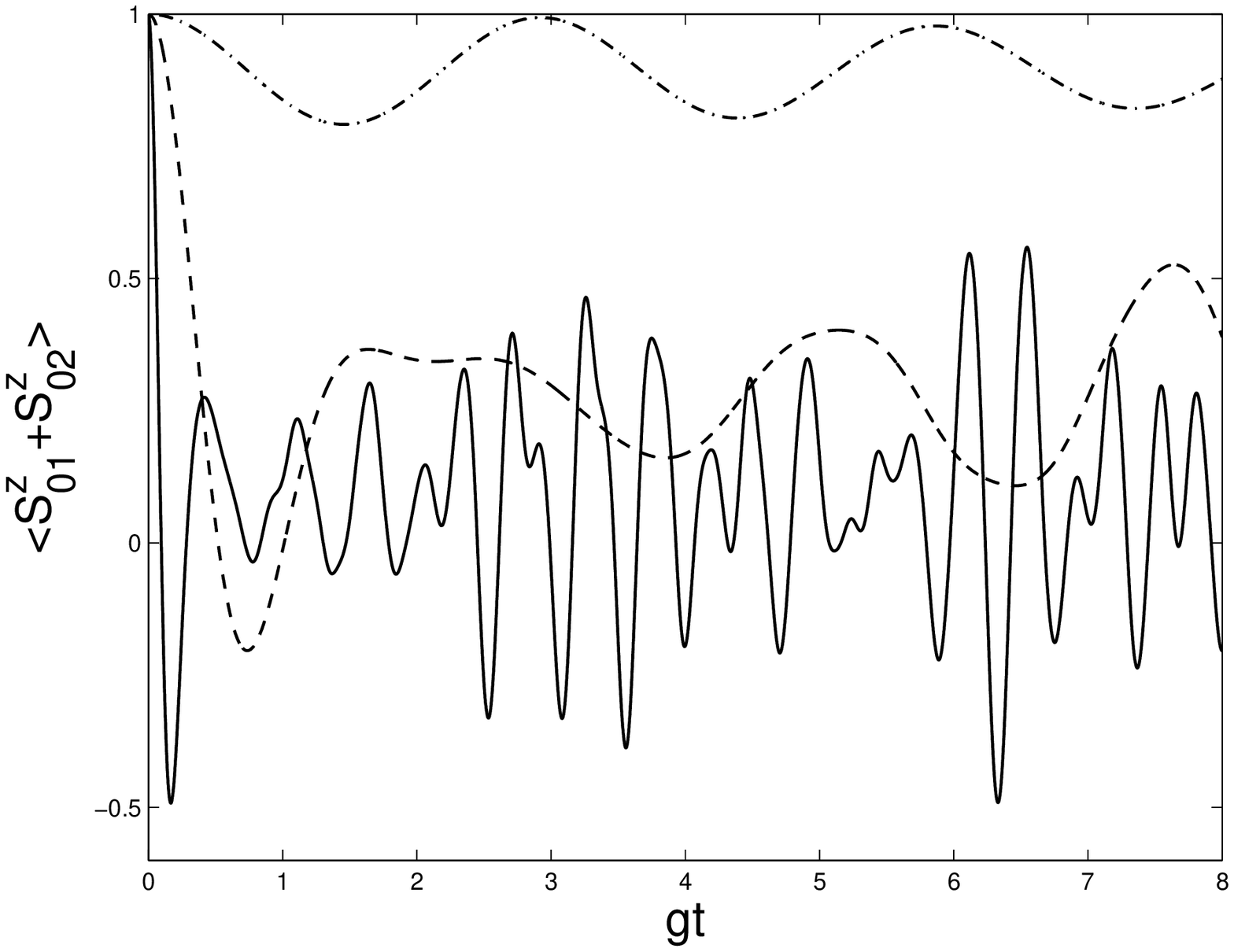}}
\caption{Time evolution of (a) purity and (b) z-component
oscillation for an initial two-qubit state of
$|\psi(0)\rangle=|11\rangle$ for different values $g_0$: $g_0=0.2g$
(dot dashed curve), $g_0=g$ (dashed curve) and $g_0=5g$ (solid
curve). Other parameters are $\mu_0=2g$, $\Omega=0g$, $T=5g$.}
\label{g011}
\end{figure}

\begin{figure}[htbp]
\centering \subfigure[$S_{vN}(t)$]{\label{T0110:entropy}
\includegraphics[width=3in]{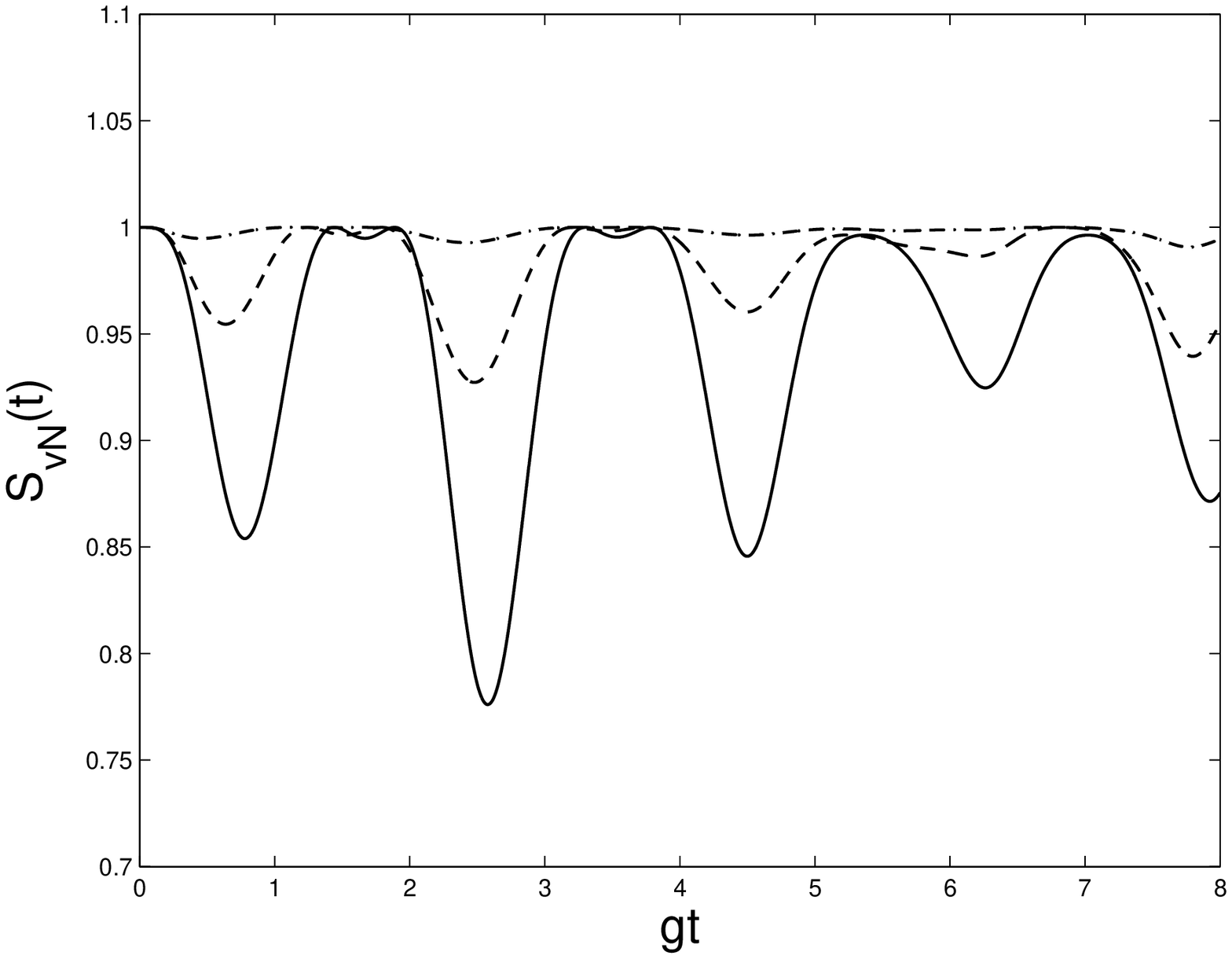}}
\subfigure[$C(t)$]{\label{T0110:Con}
\includegraphics[width=3in]{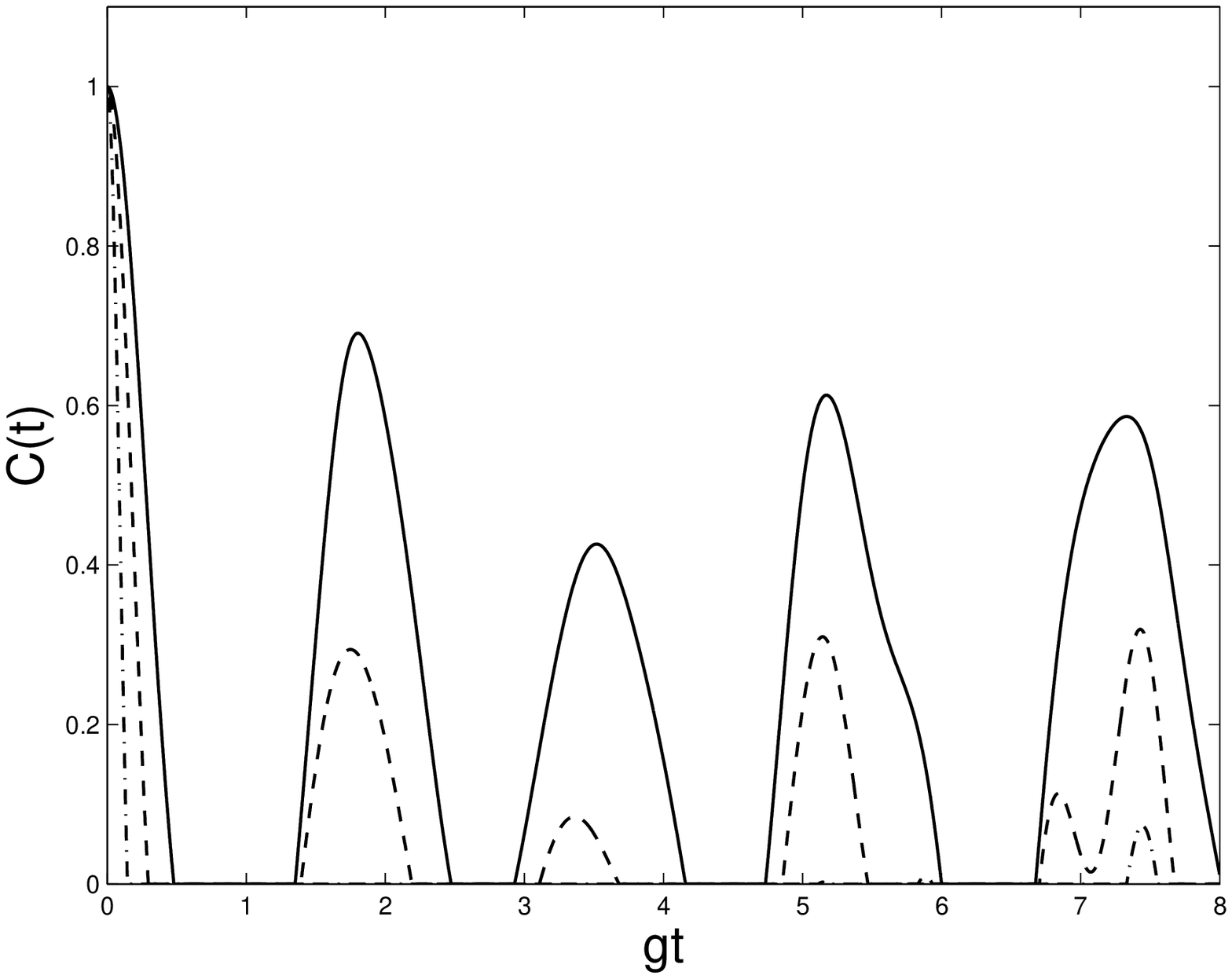}}
\caption{Time evolution of (a) von Neumann entropy and (b)
Concurrence for an initial two-qubit state of
$|\psi(0)\rangle=1/\sqrt{2}(|01\rangle+|10\rangle)$ for different
values temperature: $T=2g$ (solid curve), $T=5g$ (dashed curve) and
$T=20g$ (dot dashed curve). Other parameters are $\mu_0=2g$,
$g_0=g$, $\Omega=0g$.} \label{T0110}
\end{figure}

\begin{figure}[htbp]
\centering \subfigure[$S_{vN}(t)$]{\label{Omega0110:entropy}
\includegraphics[width=3in]{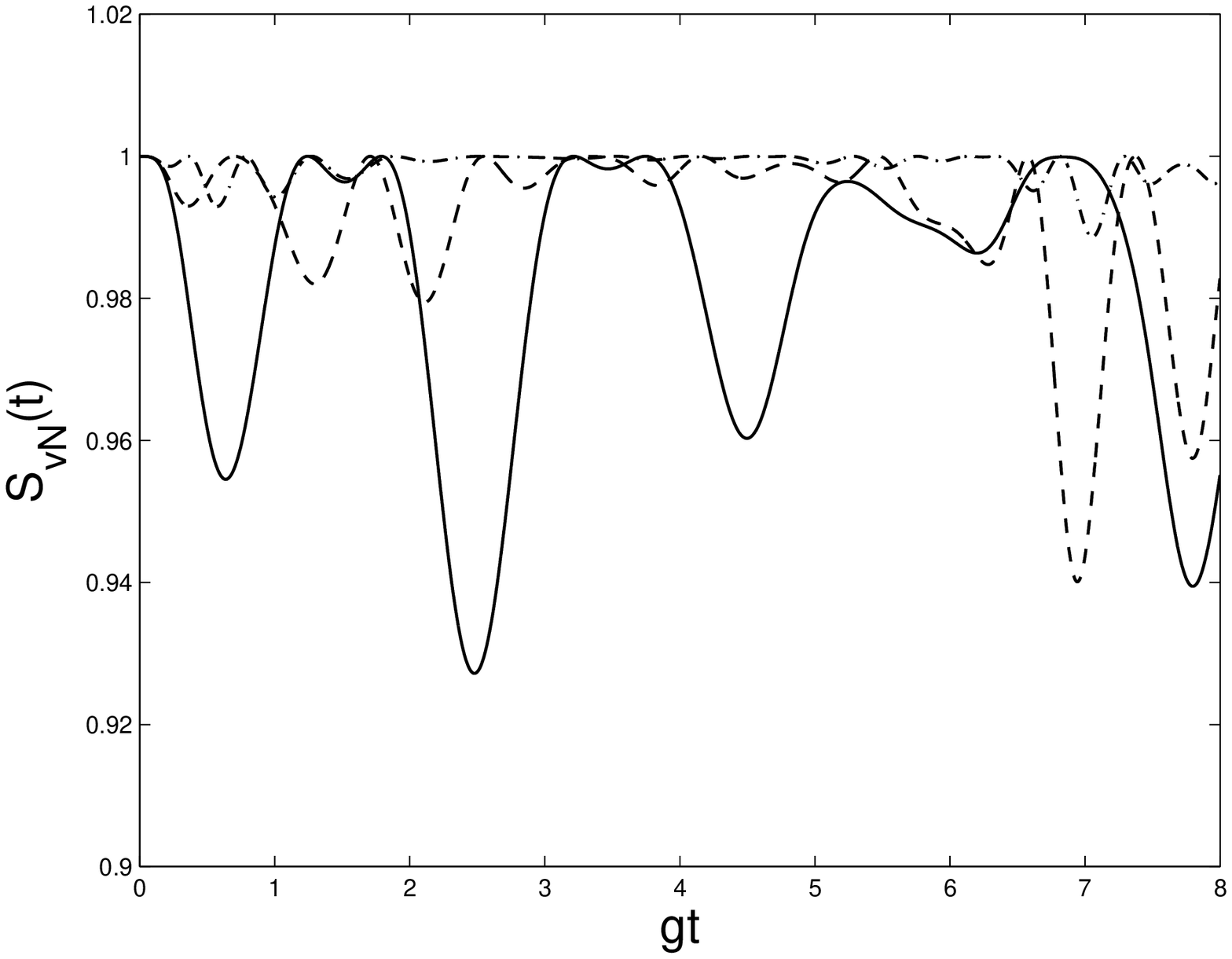}}
\subfigure[$C(t)$]{\label{Omega0110:Con}
\includegraphics[width=3in]{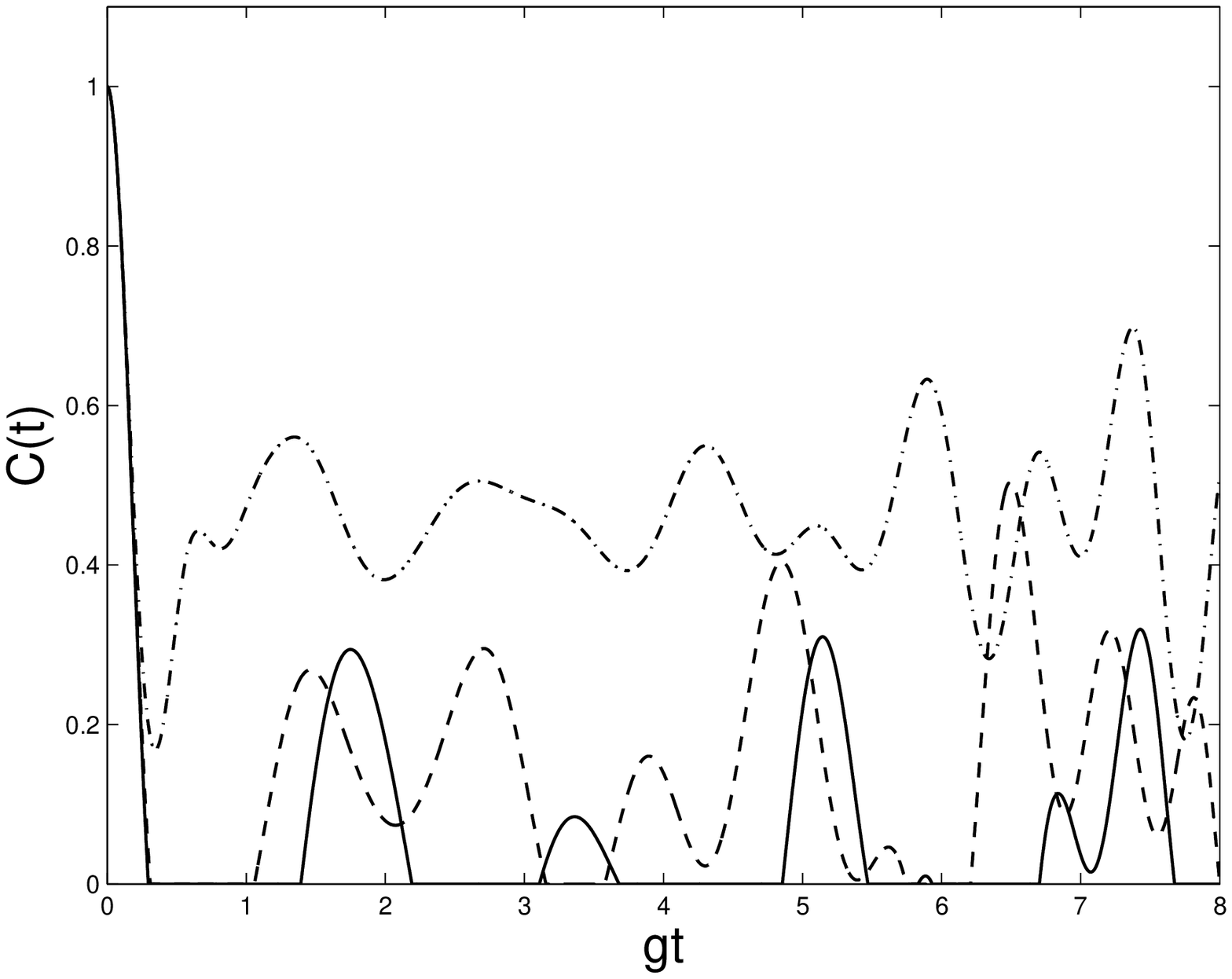}}
\caption{Time evolution of (a) von Neumann entropy and (b)
Concurrence for an initial two-qubit state of
$|\psi(0)\rangle=1/\sqrt{2}(|01\rangle+|10\rangle)$ for different
values $\Omega$: $\Omega=0g$ (solid curve), $\Omega=2g$ (dashed
curve) and $\Omega=5g$ (dot dashed curve). Other parameters are
$\mu_0=2g$, $g_0=g$, $T=5g$.} \label{Omega0110}
\end{figure}

\begin{figure}[htbp]
\centering \subfigure[$S_{vN}(t)$]{\label{g00110:entropy}
\includegraphics[width=3in]{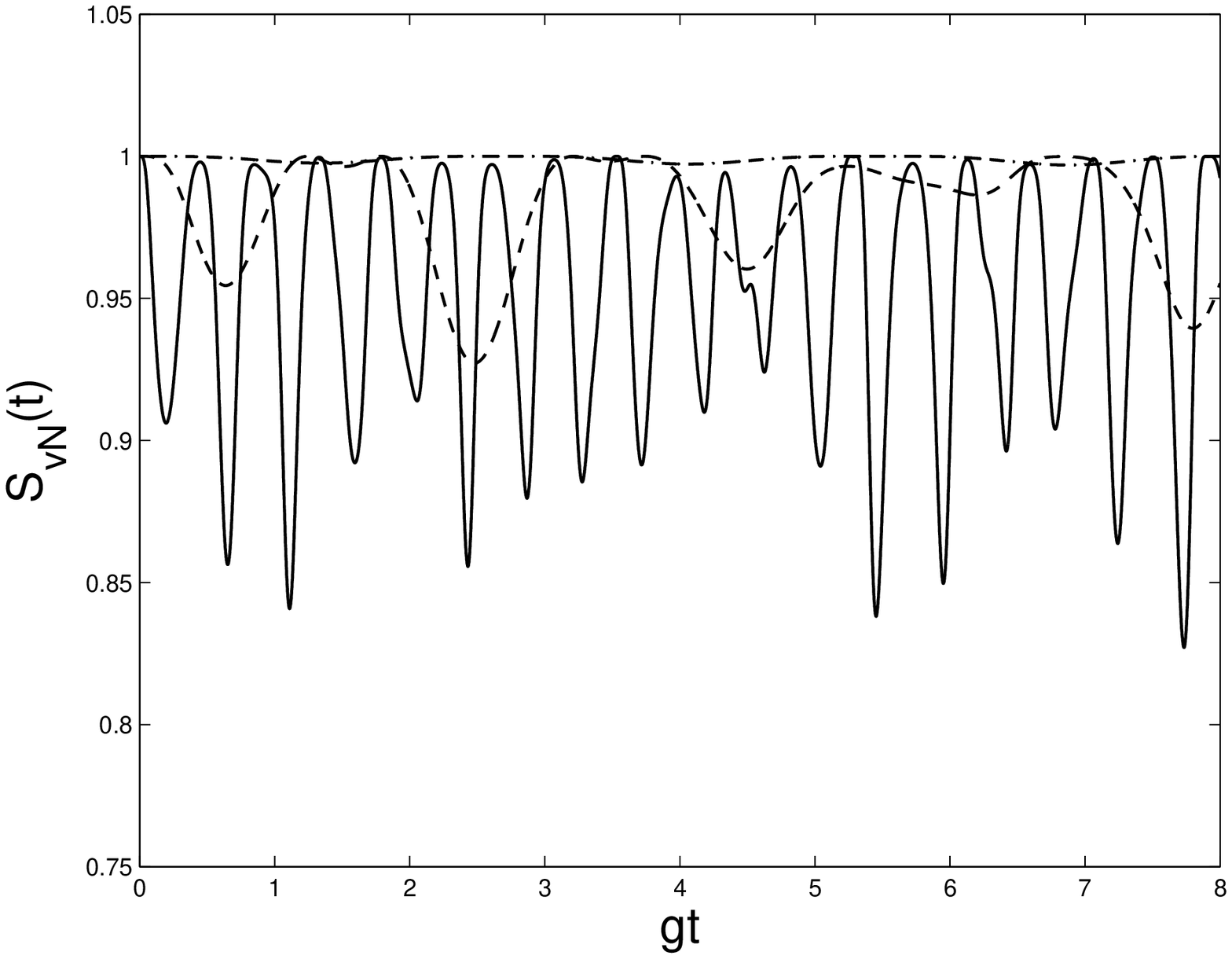}}
\subfigure[$C(t)$]{\label{g00110:Con}
\includegraphics[width=3in]{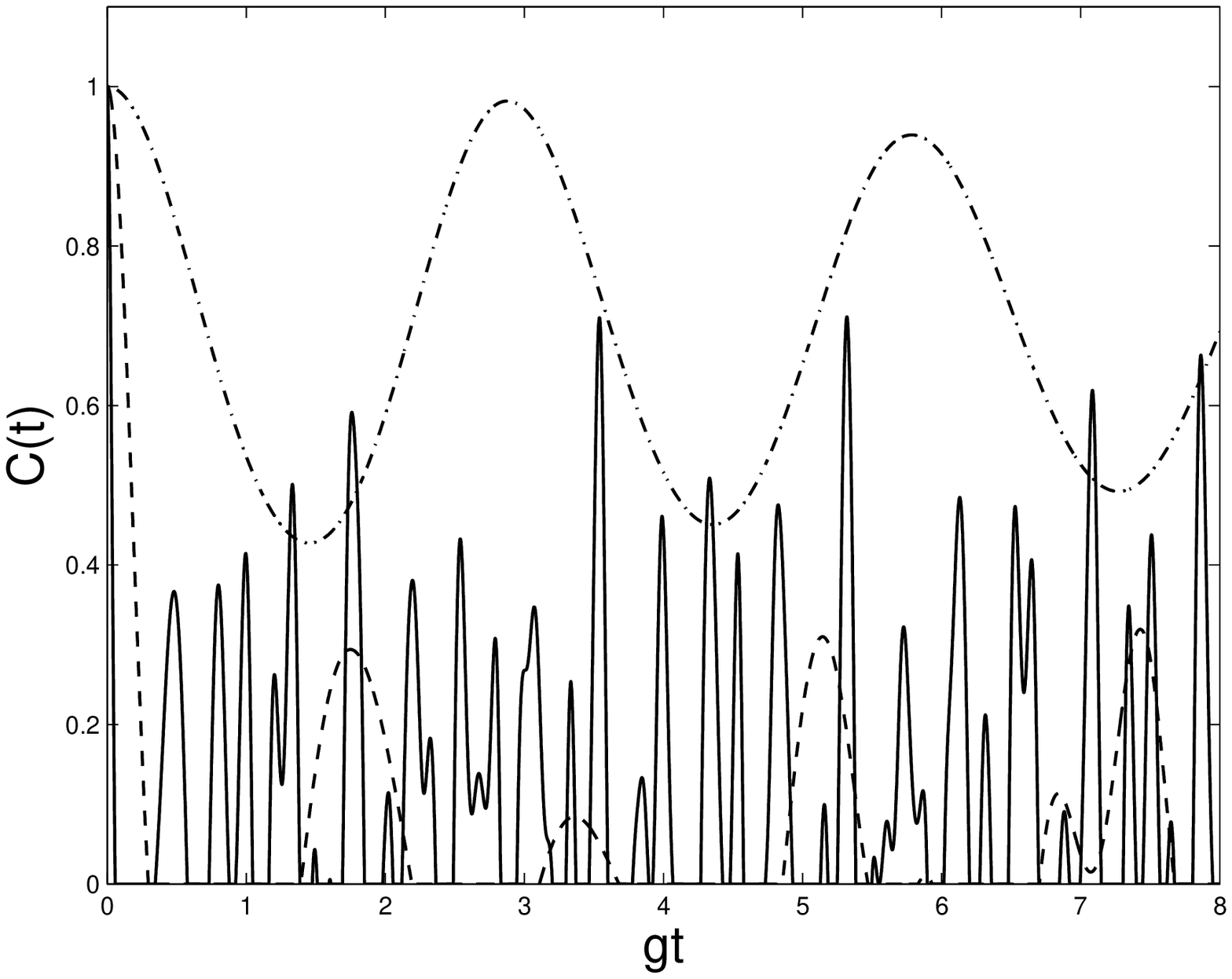}}
\caption{Time evolution of (a) von Neumann entropy and (b)
Concurrence for an initial two-qubit state of
$|\psi(0)\rangle=1/\sqrt{2}(|01\rangle+|10\rangle)$ for different
values $g_0$: $g_0=0.2g$ (dot dashed curve), $g_0=g$ (dashed curve)
and $g_0=5g$ (solid curve). Other parameters are $\mu_0=2g$,
$\Omega=0g$, $T=5g$.} \label{g00110}
\end{figure}

\begin{figure}[htbp]
\centering \subfigure[$S_{vN}(t)$]{\label{g00011:entropy}
\includegraphics[width=3in]{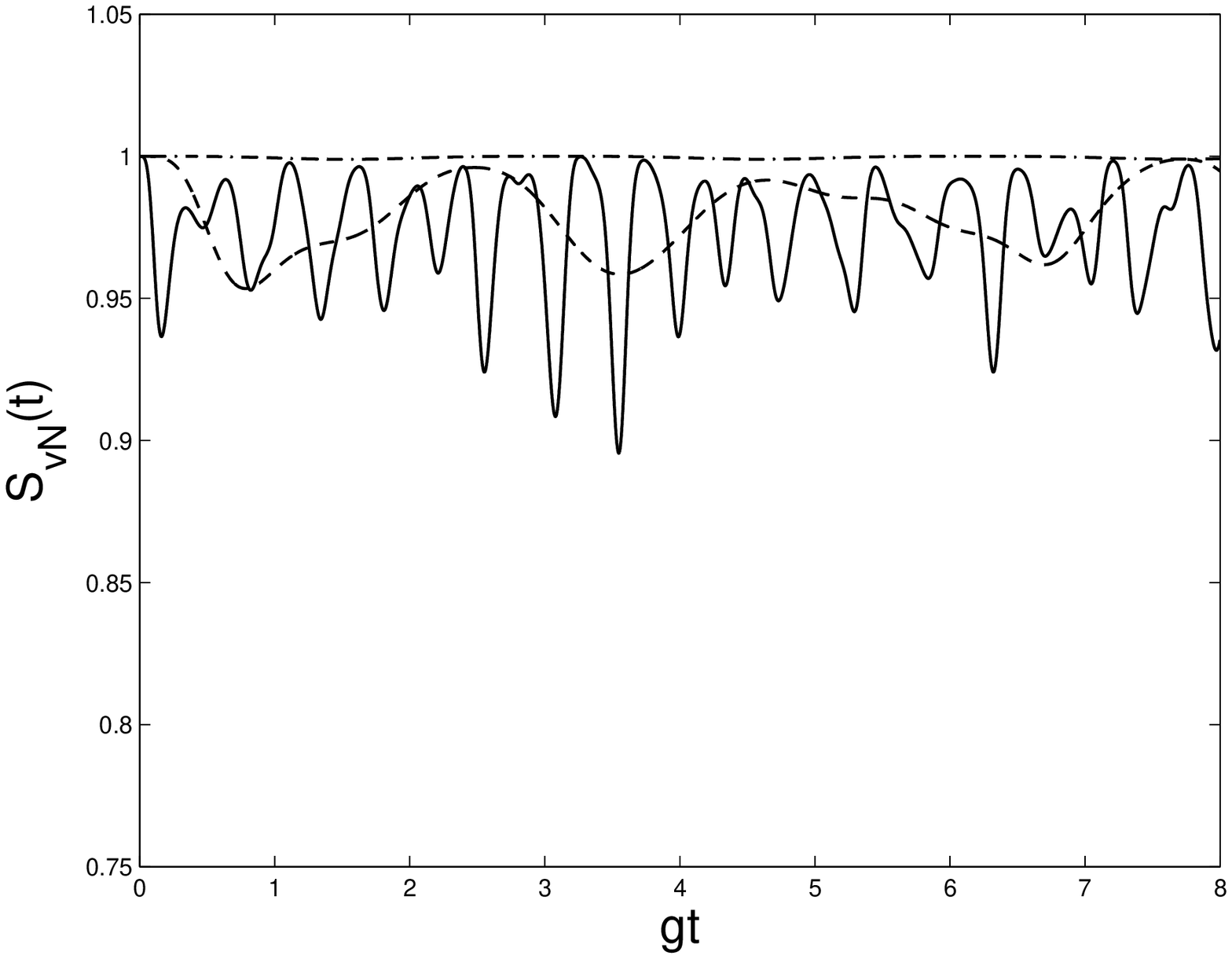}}
\subfigure[$C(t)$]{\label{g00011:Con}
\includegraphics[width=3in]{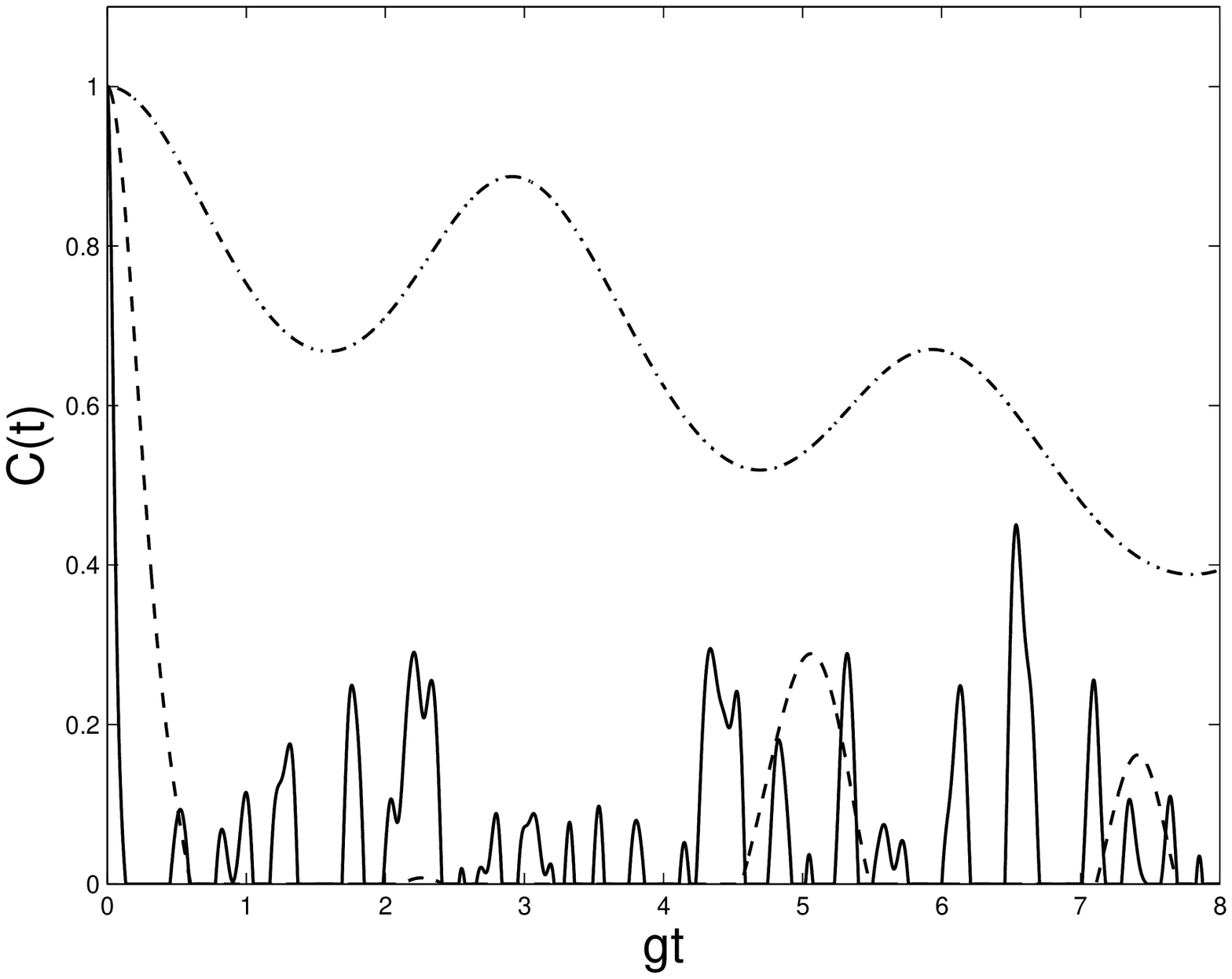}}
\caption{Time evolution of (a) von Neumann entropy and (b)
Concurrence for an initial two-qubit state of
$|\psi(0)\rangle=1/\sqrt{2}(|00\rangle+|11\rangle)$ for different
values $g_0$: $g_0=0.2g$ (dot dashed curve), $g_0=g$ (dashed curve)
and $g_0=5g$(solid curve). Other parameters are $\mu_0=2g$,
$\Omega=0g$, $T=5g$.} \label{g00011}
\end{figure}

After we obtain the reduced density matrix, we can calculate any
physical quantities of the subsystem. In the following we focus our
attention on three important physical quantities of the subsystem
which reflect the quantum entropy increase caused by decoherence,
the population inversion and the entanglement degree of the
subsystem state respectively. These quantities are (i) the time
dependence of the state purity, i.e. $P={\rm Tr}(\rho_S^2)$, which
characterizes the conservation of the purity of the subsystem
\cite{Zurek1987, Zurek1981}. If $P=1$, the subsystem is a pure
state; whereas $P=1/2^n$ (considering there are $n$ spins-1/2 in the
subsystem), it is in a completely mixed state $\rho_S=1/2^nI$ ($I$
is the identical matrix in $2^n$ dimension). (ii) the $z$-component
of the central spins, i.e. $\langle{}S^{z}_{01}+S^{z}_{02}\rangle$,
which demonstrates the population probability of the system. (iii)
the time-evolution of concurrence \cite{Wootters1, Wootters2} for
the two central spins of the open subsystem. The concurrence of the
two spin-1/2 system is an indicator of their intra entanglement,
which is defined as \cite{Wootters1}:
\begin{equation}\label{Concurrence}
C=\max\{\lambda_1-\lambda_2-\lambda_3-\lambda_4,~0\},
\end{equation}
where $\lambda_i$ are the square roots of the eigenvalues of the
product matrix
$\rho_S(\sigma^y\otimes\sigma^y)\rho^*_S(\sigma^y\otimes\sigma^y)$
in decreasing order. \\

Theoretically, our method can deal with time evolution of the
subsystem from any initial state. Here we first discuss the case of
$|\psi(0)\rangle=|11\rangle$, which means both of the two center
spins are in their excited states. The time evolution of the purity
at different temperatures is given in Fig. \ref{T11:Purity} as well
as the expecting value of the z-component in Fig. \ref{T11:Sz}. We
find that, (i) at a very low temperature, both the quantities
present a collapse and revival phenomenon, which is identical with
the two photons resonance of two two-level atoms in a cavity. In
Fig. \ref{T11:Purity}, the state of the first peak point
($gt_1=1.424$) along the solid line is
\[
\rho_S(gt_1)=0.1472|11\rangle\langle11|+0.1472|00\rangle\langle00|+
0.7056|B\rangle\langle{}B|,
\]
where $|B\rangle=1/\sqrt{2}(|01\rangle+|10\rangle)$. By the
Peres-Horodecki separability \cite{Horodecki} test, and if we make a
positive partial transposition (PPT) operation to the $1$st center
spin, we will get a new matrix:
\[
\rho_S^{T_1}(gt_1)=0.1472|11\rangle\langle11|+0.1472|00\rangle\langle00|+
0.3528(|10\rangle\langle10|+|01\rangle\langle01|+|11\rangle\langle00|+|00\rangle\langle11|),
\]
whose spectrum is $\{0.500, 0.3528, 0.3628, -0.2055\}$. It shows
explicitly that the bath can entangle the subsystem spins, although
they are separable initially. And the state of the second peak point
($gt_2=2.468$) is
\[
\rho_S(gt_2)=0.9169|11\rangle\langle11|+0.0574|00\rangle\langle00|+
0.02574|B\rangle\langle{}B|,
\]
which could be approximated as its start state
$|11\rangle\langle11|$. In fact, one can find the time evolution of
the subsystem concurrence in the condition of
$|\psi(0)\rangle=|11\rangle$ and $T=0.2g$ in Fig. \ref{Con-Pu}. It
is interesting that the bath can induce entanglement between the two
central spins periodically even without the direct coupling between
the two spins. The peak state ($t_2=4.972g$) (the solid line in Fig.
1(a))is
\[
\rho_S(t_2)=0.9992|11\rangle\langle11|+0.0008|00\rangle\langle00|+
0.00002|B\rangle\langle{}B|,
\]
which could be approximated as its initial state
$|11\rangle\langle11|$. Then the subsystem revives at this time
point and the period of the revival is about $5g$. (ii) with
increasing temperature, their oscillation magnitudes are quickly
damped by the thermal bath: For the purity at $T=10g$,
$P\rightarrow0.3$ means a much large derivation from the initial
state (to see the dot dashed curve in Fig. \ref{T11:Purity}); For
the z-component summation of the two spins, it means the
degeneration of their magnetization (to see the dot dashed curve in
Fig. \ref{T11:Sz}). Then we hold the bath at a moderate temperature
$T=5g$ to research the role of the direct coupling between the two
central spins. In Fig. \ref{Omega11:Purity}, with increasing
$\Omega$, the correlation between the two spins is strengthened so
that the leakage of the information in the open subsystem is reduced
and the collapse speed of the subsystem state slows down and
distinct revivals are observed. In Fig. \ref{Omega11:Sz}, the whole
magnetic moment oscillates around a mean value
$\langle{}S^{z}_{01}+S^{z}_{02}\rangle=0.6$ when $\Omega$ is as
large as $5g$. One can see that if the direct coupling between the
two central spins is strong enough, the two qubits can keep their
initial state from the influence of the bath. The effect of the
coupling strength between the qubit subsystem spins and bath spins
$g_0$ can be found in Fig. \ref{g011}. At a large value $g_0=5g$,
the strong interaction with the bath will quickly push the pure
state of the subsystem into a mixed state (to see the solid curve in
Fig. \ref{g011:Purity}); on the contrary, at a small value
$g_0=0.2g$, the initial state can be conserved to a great extent,
$P>0.68$, as the dot dashed curve shows in Fig. \ref{g011:Sz}.\\

Then we show the results of the temporal evolution for the quantum
entropy and concurrence from a most entangled state (one of the
well-known Bell states)
$|\psi(0)\rangle=1/\sqrt{2}(|01\rangle+|10\rangle)$ under the
influence of the bath. As is known, the entropy of either particle
inside a Bell states is $1$, which is certainly in a most mixed
state. On one hand, in Fig. \ref{T0110:entropy}, at any finite
temperature, the entropy will return $1$ after some fluctuation; on
the other hand, in Fig. \ref{T0110:Con}, the entanglement degree
between the two coupled spins will quickly collapse to zero and make
some fluctuations and revivals accidently. The magnitudes of these
fluctuations decrease with increasing temperature in both figures.
The damping speed of concurrence evidently increases with
temperature in Fig. \ref{T0110:Con} and when $T=20g$, it shows a
sudden death in a much short time and a tiny revival in a long time
scale. In Fig. \ref{Omega0110:entropy}, the effect of subsystem
inner-coupling $\Omega$ on the entropy is so weak that the biggest
fluctuation of that is less than $0.08$ and when $\Omega=5g$, this
effect can be neglected. Yet, in Fig. \ref{Omega0110:Con}, the
concurrence can be restored to some extent with a comparatively
larger $\Omega$: when $\Omega=5g$, there is no sudden death of the
entanglement between the subsystem spins and the concurrence
oscillates about $C=0.4$. From the above results, we find that the
von-Neumann entropy is not a very good measure of this subsystem
(consisted of two or more atoms) state as concurrence. In Fig.
\ref{g00110}, with decreasing interaction between the open subsystem
and the bath $g_0$, the influence of the bath spins on the subsystem
is reduced. As is depicted in Fig. \ref{g00110:Con}, at $g_0=0.2g$,
the dot dashed line oscillates around the $C=0.7$. This is evident
that if we can keep the coupling between the subsystem and the bath
as weak as possible, then the entanglement of subsystem can be
conserved as
much as possible.\\

With the combination of the technique of Holstein-Primakoff operator
and our polynomial numerical scheme, we have thoroughly discussed
the quantum dynamics of two central spins in a spin star lattice
within XY interaction from different kinds of initial states. In
principle, all the physical quantities can be calculated as the
above ones. For the product states, it is noticed that the bath can
help to entangle the two separable spins; For the Bell states, it is
valuable that $g_0$, the interaction between the subsystem and
environment, can be adjusted to slower the collapse of the most
entangled state. However, we also find that the behavior of the
reduced density matrix for the open subsystem relies on different
initial state although they are all of the Bell states. By the
comparison of Fig. \ref{g00110} with Fig. \ref{g00011}, it is shown:
in the former figure, when $g_0$ is small, the concurrence can be
maintained for a long time; while in the latter one, when $g_0$ is
enhanced, it is quickly damped.

\section{Conclusion}\label{conclusion}

We have studied the quantum dynamics of the purity, z-component
summation and concurrence of two coupled spin qubits in a apin bath
via Heisenberg $XY$ interaction.  A novel numerical polynomial
scheme is used in the calculation for the reduced density matrix of
the central qubits after the model Hamiltonian has been performed by
the well-known Holstein-Primakoff transformation. The procedure
avoids the difficulty in Ref. \cite{Yuan} that the initial state is
limited by the analytic derivation. The time evolution of different
types of initial states, either product states or Bell ones, is
obtained. Although the subsystem is initially prepared in the
product state ($|11\rangle$) , it turns out that the bath can induce
entanglement on the subsystem spins by the Peres-Horodecki
separability test on the state $\rho_S^{T_1}(gt_1)$. On the other
hand,  the effect of different types of coupling on the entanglement
of the system is studied. Generally, the coupling to the environment
reduces the initial state entanglement. When  the interaction
between the two spins is large,   they are protected strongly
against the environment. Thus, it is found that the dynamics of the
subsystem depends sensitively on the initial state, the bath
temperature, the inner-coupling between the two central spins and
the interaction between the subsystem and the environment. Besides,
our numerical scheme is simple and independent of the initial state,
which can be easily applied to the studies of some other kinds of
open quantum systems.

\begin{acknowledgments}
We would like to acknowledge the support from the China National
Natural Science Foundation.
\end{acknowledgments}

\end{document}